# Designing graphene/hexagonal boron nitride superlattice monolayer with high thermoelectric performance

Zizhen Zhou, Huijun Liu[*], Dengdong Fan and Guohua Cao

*Key Laboratory of Artificial Micro- and Nano-Structures of Ministry of Education and School of Physics and Technology, Wuhan University, Wuhan 430072, China*

**Abstract**

We design a hybrid graphene/hexagonal boron nitride superlattice monolayer and investigate its thermoelectric properties using density functional theory and Boltzmann transport equations with the relaxation time accurately treated by electron-phonon coupling calculations. Compared with that of pristine graphene, the lattice thermal conductivity of the superlattice structure is more than two orders of magnitude lower due to the enhanced three-phonon scattering process originated from the mixed-bond characteristics. Besides, the coexistence of light and heavy bands around the Fermi level leads to an ultrahigh power factor along the zigzag direction, where the highest *ZT* value of ~2.5 can be achieved for the *n*-type system at 1100 K. Moreover, it is noted that the carrier transport near the valance band minimum is almost entirely contributed by the graphene part of the superlattice. As a consequence, the thermoelectric performance of *p*-type system can be enhanced to be comparable with that of *n*-type one by appropriate substitution of nitrogen atom with phosphorus, which can suppress the lattice thermal conductivity but nearly have no influence on the hole transport.

## 1. Introduction

As an environmentally friendly energy conversion technique, thermoelectric (TE) power generation has attracted renewed interest since it can harvest waste heat and directly convert it into electricity. The TE performance of a material is usually determined by the figure-of-merit $ZT = S^2\sigma T/(\kappa_e + \kappa_l)$, where $S$ is the Seebeck coefficient, $\sigma$ is the electrical conductivity, and $T$ is the absolute temperature, with $\kappa_e$ and $\kappa_l$ representing the electronic and lattice thermal conductivity, respectively. Good TE material should combine low thermal conductivity with high electrical conductivity and large Seebeck coefficient, which is usually difficult to be realized as a result of strong coupling between these transport coefficients. Compared with the bulk structures, the low-dimensional systems have been considered to be superior in achieving efficient TE conversions [1−3]. As a typical single-layer material, graphene has many extraordinary electrical and optical properties with great potential in variety of applications [4−6]. In spite of the intrinsic weaknesses such as excessive thermal conductivity [7] and gapless energy band [8], the TE application of graphene has still attracted some interest due to its ultrahigh carrier mobility [9] and good mechanical properties [10].

---

[*] Author to whom correspondence should be addressed. Electronic mail: phlhj@whu.edu.cn

During the past decades, many effective methods have been suggested to improve the TE performance of graphene, such as inducing antidot lattice [11], constructing stub structure of nanoribbons [12], and utilizing surface functionalization [13]. Among various strategies, designing superlattice structures, which was demonstrated to be efficient in realizing higher *ZT* values [14], has attracted much attention, especially for the in-plane heterostructures of graphene and hexagonal boron nitride (h-BN) [15−20]. Experimentally, a novel nanomaterial consisting of graphene and h-BN was synthesized successfully using the thermal catalytic chemical vapor deposition method [21], which suggested the possibility to form superlattice monolayer [22] and stimulated new strategies of band gap engineering [23]. Subsequently, several theoretical works demonstrated that tunable electronic band gaps emerge in the in-plane graphene/h-BN heterostructures made of adjacent armchair ribbons [24−26], where a very high Seebeck coefficient could be achieved [27]. Besides, the phonon transport in such types of superlattice monolayers was also investigated by molecular dynamics simulations [28−31], where the lattice thermal conductivities are obviously lower than those of the pristine graphene and could be favorable to achieve high TE performance. Using the nonequilibrium Green's function approach, Yang *et al.* [32] showed that the hybrid graphene/h-BN nanoribbons indeed exhibit much higher room temperature *ZT* values compared with those of the components. So far, the TE performance of graphene/h-BN superlattice monolayers reported in previous studies [33−35] were still obviously lower than that of the state-of-the-art bulk material such as SnSe [36]. It is thus natural to ask if higher *ZT* with comparable *p*- and *n*-type values could be realized in the superlattice monolayer consisting of the graphene and h-BN nanoribbons.

In this work, the electronic, phonon, and TE transport properties of graphene/h-BN superlattice monolayer are investigated by combining first-principles calculations and Boltzmann transport theory. It is found that the TE performance exhibits strong anisotropy and obvious temperature dependence, with the highest *n*-type *ZT* value of ~2.5 obtained along the zigzag direction at 1100 K. Furthermore, we show that the *p*-type TE performance can be enhanced to be comparable with that of the *n*-type system via moderate substitution of the phosphorus with the nitrogen atoms.

## 2. Computational details

The lattice thermal conductivity of the graphene/h-BN superlattice monolayer is computed by solving the phonon Boltzmann transport equation, as implemented in the so-called ShengBTE package [37]. The harmonic and anharmonic properties are investigated by density functional theory (DFT) [38,39] combined with the finite displacement method using a $11\times3\times1$ supercell. The eighth nearest neighbors are included for the third-order interactions to ensure convergence. The DFT calculations are implemented in the so-called Vienna *ab-initio* simulation package (VASP) [40], and the second- and third-order interatomic force constants are computed via the so-called PHONOPY program [41] and the THIRDORDER.PY script [37], respectively. A fine $151\times51\times1$ *q*-mesh is adopted to yield converged lattice thermal conductivity.

The electronic properties are calculated within the framework of DFT, as performed in the QUANTUM ESPRESSO package [42]. The norm-conserving scalar-relativistic pseudopotential is used to describe the core-valence interaction [43], and the exchange-correlation functional is in the form of Perdew-Burke-Ernzerhof (PBE) with the generalized gradient approximation (GGA) [44]. The superlattice structure is modeled by adopting a rectangular cell with the vacuum distance of 30 Å to eliminate the interactions between the periodic images. Using a dense $65\times21\times1$ $\boldsymbol{k}$-mesh and a kinetic energy cutoff of 700 eV, the system is fully relaxed until the magnitude of the forces acting on all the atoms becomes less than $10^{-3}$ eV Å$^{-1}$. To obtain accurate electronic transport coefficients, the band structure of the superlattice monolayer is calculated by adopting hybrid functional in the form of Heyd-Scuseria-Ernzerhof (HSE) [45], which has been successfully used to predict the band gaps of the graphene-like BN monolayer [46]. Within the framework of Boltzmann transport theory [47], the transport coefficients $S$, $\sigma$, and $\kappa_e$ are calculated with the $\boldsymbol{k}$-resolved carrier relaxation time ($\tau_c$) obtained from a complete electron-phonon coupling (EPC) calculation [48]. The convergence is carefully checked by using coarse grids of $10\times10\times1$ $\boldsymbol{k}$-points and $5\times5\times1$ $\boldsymbol{q}$-points, and dense meshes of $200\times200\times1$ $\boldsymbol{k}$-points and $100\times100\times1$ $\boldsymbol{q}$-points obtained from the Wannier interpolation techniques [49]. To have a better comparison, the transport coefficients $\sigma$, $\kappa_e$, and $\kappa_l$, as well as the carrier concentration, are all renormalized with respect to the interlayer distance of graphite.

## 3. Results and discussion

As can be seen from Fig. 1(a), the investigated superlattice monolayer is consist of alternating arrangement of zigzag-edged graphene and h-BN nanoribbons with perfect periodicity along the armchair direction. The structure contains eight atoms ($C_4B_2N_2$) in a rectangular primitive cell as framed by the black lines, where the lattice parameters along the zigzag and armchair directions are calculated to be 2.483 Å and 8.689 Å, respectively. There are eight different covalent bonds in the crystal structure, with the lengths of 1.424 Å, 1.426 Å, 1.445 Å, 1.390 Å, 1.530 Å, 1.451 Å, 1.448 Å, and 1.429 Å for the C1−C3, C2−C4, C2−C3, C1−N2, C4−B1, B1−N1, B2−N2, and N1−B2 chains, respectively. The mixed covalent bonds with different length suggest relatively stronger phonon scattering and thus lower lattice thermal conductivity, which is beneficial to achieve high TE performance as will be discussed later.

Fig. 1(b) shows the phonon dispersion relations of the graphene/h-BN superlattice calculated by density functional perturbation theory [50], where there is no imaginary frequency guaranteeing the dynamic stability of the system. Similar to that of the h-BN [51], the superlattice monolayer exhibits exceptional thermal stability since the structure remains unchanged in the *ab-initio* molecular dynamics simulation at 1500 K (see Fig. S1 of the Supplemental Material). Compared with that of graphene, we see more pronounced hybrid characteristic of acoustic and optical branches in the phonon dispersion relations of the superlattice monolayer, which can be attributed to the vibrational mismatch between different atoms [52]. Indeed, we find distinct phonon modes of C, B, and N atoms at the $\Gamma$ point near the frequency of 200 cm$^{-1}$ (not shown here) caused by the mixed-bond characteristics [53], leading to enhanced anharmonic phonon scattering rates and thus lower lattice thermal conductivity [54]. By solving the phonon Boltzmann transport equation, the temperature-dependent $\kappa_l$ of the graphene/h-BN superlattice can be obtained. As plotted in Fig. 1(c), the $\kappa_l$ exhibit obvious anisotropy rooted from the different bonding configurations along the zigzag and armchair directions. To confirm the reliability of our approach, we have done additional calculation on the lattice thermal conductivity of graphene. As shown in the inset of Fig. 1(c), we find that the room temperature value is 5660 W/mK which agrees well with that reported previously [7]. In the whole temperature region, it is amazing to find that the $\kappa_l$ of the superlattice are more than two orders of magnitude lower than those of the graphene, with the values at 300 K to be 19.9 and 11.6 W/mK along the zigzag and armchair direction, respectively. Note that these two systems exhibit similar group velocities, and the distinct lattice thermal conductivities should thus be attributed to their different phonon relaxation time ($\tau_p$). Fig. 1(d) and its inset plot the room temperature $\tau_p$ of the superlattice and graphene, respectively. It is found that the relaxation times of the acoustic phonons in the superlattice are nearly two orders of magnitude lower than those in the graphene, which contribute more than 95% of the lattice thermal conductivity owing to the selection rules [55]. Such an observation suggests that the heat transport contributed by the acoustic modes is dramatically limited in the superlattice monolayer and should be responsible for its much lower lattice thermal conductivity. Besides, the optical phonon relaxation time of the superlattice monolayer is also at least an order of magnitude lower compared with that of graphene. As a consequence, the extremely lower $\kappa_l$ in the superlattice can be attributed to the sharply reduced $\tau_p$ due to more pronounced hybrid characteristic of acoustic and optical branches, which should be traced back to the mixed covalent bonds as discussed above.

We now move to the discussion of the electronic band structures of the graphene/h-BN superlattice monolayer, which are computed by using HSE functional. A careful search in the whole Brillouin zone find that both the conduction band minimum (CBM) and the valance band maximum (VBM) are located at the **k**-point of (0.437, 0.500, 0.000). The corresponding band gap is 1.48 eV, which is relatively larger than previous result [27] calculated by using PBE functional. Fig. 2(a) plots the orbital-decomposed band structures of the superlattice, which are dominated by the $p_z$-orbitals of each atom in the energy window from −4 to 4 eV. For the top valence band (bottom conduction band), the weak dispersions near the Fermi level are mainly contributed by the $p_z$-orbitals of B and C (N and C) atoms. On the other hand, the strong dispersions at the band edges are almost entirely occupied by the $p_z$-orbitals of C atoms, indicating that the carrier transport may be dominated by the graphene of the superlattice monolayer. The coexistence of such light and heavy bands around the Fermi level is very beneficial for achieving high TE performance [56−58], since the flat bands provide high electronic density of state (DOS) and induce a large Seebeck coefficient, while the dispersive bands facilitate carrier transport owing to the low effective mass. To have a better understanding, we calculate the energy dispersion relations of the top valence band and bottom conduction band in the whole Brillouin zone as shown in Fig. 2(b). Due to the mirror symmetry of the lattice structure and the energy bands, there are additional valance and conduction band extremum at the

*k*-point of (0.563, 0.500, 0.000), which exhibit entirely identical energies with those of the VBM and CBM, respectively. Hence, the off-symmetry VBM and CBM both show a band degeneracy of four, which could directly enhance the electrical conductivity by increasing carrier concentration for a given Fermi level without reducing the Seebeck coefficient owing to the band misalignment [59]. Besides, we see clearly strong and weak energy dispersions along the zigzag and armchair directions, respectively. The corresponding effective mass is calculated to be 0.179 (0.159) $m_e$ and 4.243 (3.704) $m_e$ around the VBM (CBM). Such unique band structure means that the electrical conductivity along the zigzag direction should be significantly larger than that along the armchair direction, which may lead to a higher power factor [58].

Fig. 2(c) shows the *k*-resolved carrier relaxation time of the graphene/h-BN superlattice monolayer at 300 and 1100 K, which are obtained from a complete EPC calculation. Compared with that of the deformation potential (DP) theory [60] where only the scattering of the acoustic phonons is considered, we find that the EPC calculated relaxation time (344 fs) of the superlattice near the CBM at 300 K is obviously lower. It is reasonable since more optical phonons are populated owing to the mixed characteristics of covalent bonds, which can't be ignored when estimating the carrier scattering rate. Besides, it can be found that the $\tau_c$ of the *n*-type system near the Fermi level is somewhat larger than that of the *p*-type system, which is caused by the constricted phase space and thus limited carrier scattering process around the CBM originated from the smaller band effective mass [61]. Inserting the energy-dependent relaxation time, the electronic properties ($S$, $\sigma$, and $\kappa_e$) of the superlattice monolayer can be obtained from the Boltzmann transport theory. For both the *p*- and *n*-type systems, we observe similar room temperature Seebeck coefficient (absolute value) but much larger electrical conductivity along the zigzag direction when compared with those along the armchair direction (see Fig. S2 of the Supplemental Material). We thus focus on the zigzag direction in the following discussions. On the other hand, the Seebeck coefficient of the *p*-type system is slightly higher than that of the *n*-type system ascribed to the bigger hole effective mass and thus larger DOS, whereas the $\sigma$ is relatively lower caused by the smaller relaxation time. Benefited from the mixture of flat and dispersive bands mentioned above, an ultrahigh room temperature power factor of 0.146 W/mK$^2$ can be obtained along the zigzag direction for the *n*-type system, which is significantly larger than that of good TE materials such as SnSe [36]. As a consequence, excellent TE performance can be realized in the graphene/h-BN superlattice monolayer. For both the *p*- and *n*-type systems, the optimized TE performance appear at a high temperature of 1100 K (see Fig. S3 of the Supplemental Material) owing to relatively larger band gap. As can be seen from Fig. 2(d), a highest *ZT* value of ~2.5 can be achieved at 1100 K along the zigzag direction at the optimized electron concentration of $3.9 \times 10^{20}$ cm$^{-3}$. The corresponding transport coefficients are summarized in Table S1 of the Supplemental Material. It should be noted that the TE module always needs both *n*- and *p*-type legs with comparable conversion efficiency to form the *p*–*n* junction. However, the *ZT* value of the *p*-type graphene/h-BN superlattice monolayer are obviously smaller than that of the *n*-type system becaues of the lower power factor as shown in Fig. 2(d). Hence, it is quite necessary to enhance the TE performance of the *p*-type system by appropriate strategies such as chemical doping or strain engineering, which requires a deep understanding of the carrier transport mechanism.

Fig. 3(a) and 3(b) show the band decomposed charge density of the graphene/h-BN superlattice monolayer around the CBM and VBM, respectively. For both cases, we can clearly see that the charge density is extended along the zigzag direction but localized along the armchair direction, which is consistent with the strongly anisotropic electrical conductivity discussed above. Besides, it is interesting to find that the C1−C3 and B2−N2 bonds dominate the electron transport near the CBM, whereas the hole transport around the VBM is mainly contributed by the C1−C3 and C2−C4 chains. It is reasonable to expect that enhanced *p*-type TE performance could be realized by introducing vacancies or impurities in the B/N sites, which should dramatically suppress the lattice thermal conductivity without affecting the hole transport. To confirm this point, we consider the case where the atoms at N2 site are substituted by the P atoms with an atomic concentration of ~2%, as circled in Fig. 3(b). A 3×2×1 supercell with nominal formula of $C_{24}B_{12}N_{11}P$ is constructed to model such a doped system. As shown in Fig. S4 of the Supplemental Material, the phonon spectrum of $C_{24}B_{12}N_{11}P$ exhibits no imaginary frequency ensuring the dynamic stability of the system. Compared with those of the pristine system, it can be found that the three acoustic modes (LA, TA, ZA) are remarkably softened owing to the distortion of chemical environments around the P atoms. The highest frequencies of the LA, TA, and ZA branches are reduced from 571 cm$^{-1}$, 558 cm$^{-1}$, and 290 cm$^{-1}$ to 400 cm$^{-1}$, 328 cm$^{-1}$, and 73 cm$^{-1}$, respectively, leading to smaller acoustic phonon group velocity and thus lower lattice thermal conductivity. Indeed, the $\kappa_l$ along the zigzag direction plotted in Fig. 3(c) is decreased by nearly 30% upon the substitution of P, which is also found in many other materials [62−65]. Fig. 3(d) plots the power factor and $ZT$ value of the P-substituted graphene/h-BN superlattice monolayer as a function of carrier concentration along the zigzag direction at 1100 K. As expected, the *p*-type power factor almost keeps unchanged by substitution. This is reasonable since the flat bands around the VBM which largely determine the Seebeck coefficient are dominated by the C and B atoms. Meanwhile, the hole transport is dominated by the C1−C3 and C2−C4 bonds coinciding with the fact that the strong dispersions near the VBM are almost entirely contributed by the C atoms as discussed above. However, this is not the case for the *n*-type system, where the $S^2\sigma$ is obviously reduced owing to the lower electrical conductivity. Our additional calculations indicate that the electron band effective mass of the P-substituted superlattice (0.201 $m_e$) is significantly larger than that of the pristine system along the zigzag direction, which leads to lower relaxation time and thus decreased electrical conductivity. As the lower power factor is largely compensated by the reduced lattice thermal conductivity, the maximum *n*-type $ZT$ value (~2.6, along the zigzag direction) of the P-substituted superlattice is similar to that of the pristine system at 1100 K. On the contrary, it is interesting to find that the TE performance of the *p*-type substituted superlattice is greatly enhanced ascribed to the decreased $\kappa_l$ and almost unchanged $S^2\sigma$. As a consequence, comparable *ZT* values (~2.6) for both the *p*- and *n*-type systems can be realized in the superlattice monolayer along the zigzag direction at 1100 K, which is highly desirable as the thermoelectric devices.

## 4. Summary

We demonstrate by first-principles study that the graphene/h-BN superlattice monolayer could achieve a record high *n*-type *ZT* value of ~2.5 along the zigzag direction at 1100 K, and comparable *p*-type TE performance is realized by substituting the P atoms in the N2 sites with an atomic concentration of ~2%. To experimentally check our strong predictions, one needs to first fabricate the superlattice monolayer. In this regard, the recent discovery of layered electride [66,67] provides an attractive family of substrate for the epitaxial growth of two-dimensional materials with particular structures [68]. Especially, it shows significant advantage over the metal substrate and offers a new way to synthesize the in-plane superlattice. In a word, our theoretical work suggests that the superlattice monolayer consisting of light, earth-abundant, and environment-friendly elements can be designed as perfect TE modules with comparable *p*- and *n*-type energy conversion efficiency.

**Acknowledgements**

We thank financial support from the National Natural Science Foundation (Grant Nos. 51772220 and 11574236). The numerical calculations in this work have been done on the platform in the Supercomputing Center of Wuhan University.

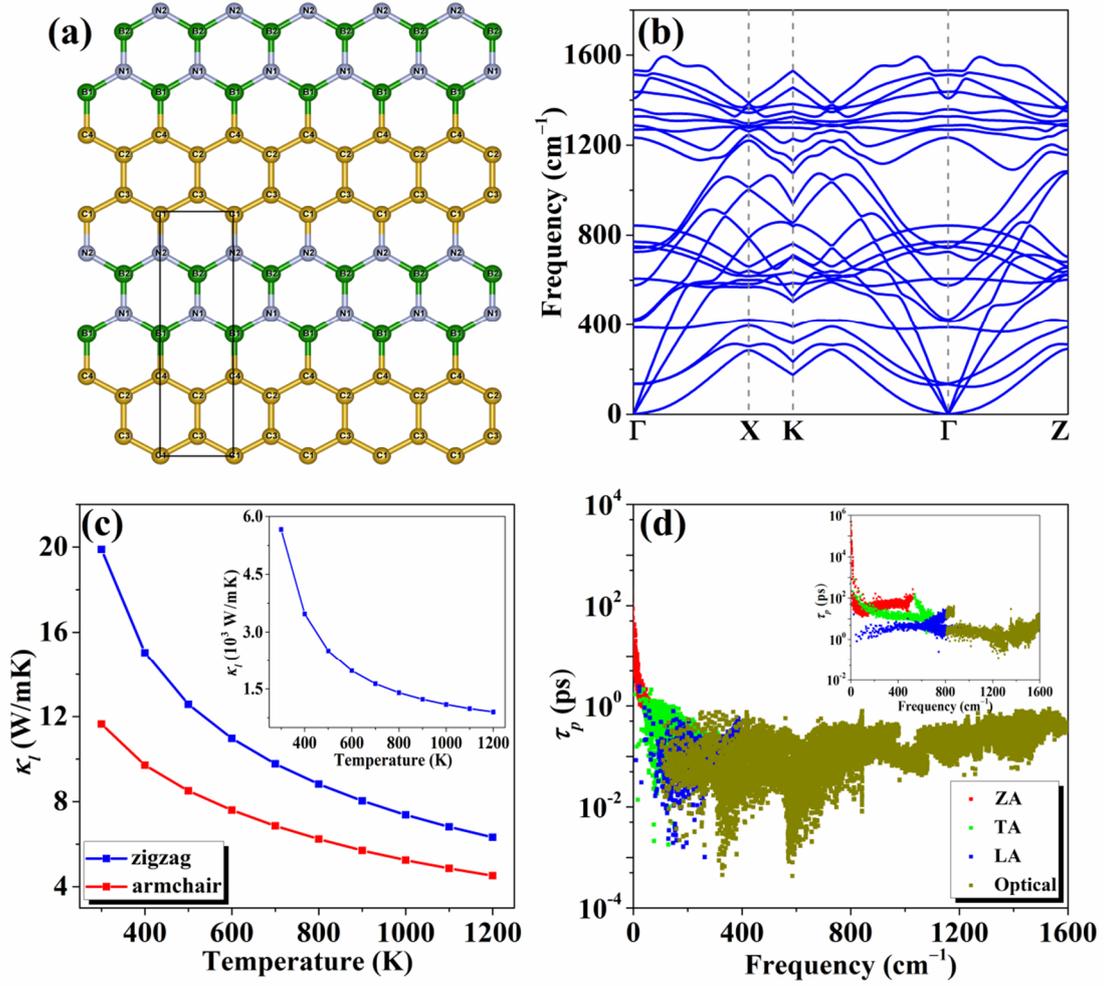

**Figure 1.** (a) Ball-and-stick model of the graphene/h-BN superlattice monolayer where the rectangle indicates the primitive cell. (b) The corresponding phonon dispersion relations. (c) The temperature-dependent lattice thermal conductivity. (d) The room temperature phonon relaxation time as a function of frequency. The insets in (c) and (d) are the lattice thermal conductivity and the room temperature phonon relaxation time of the graphene, respectively.

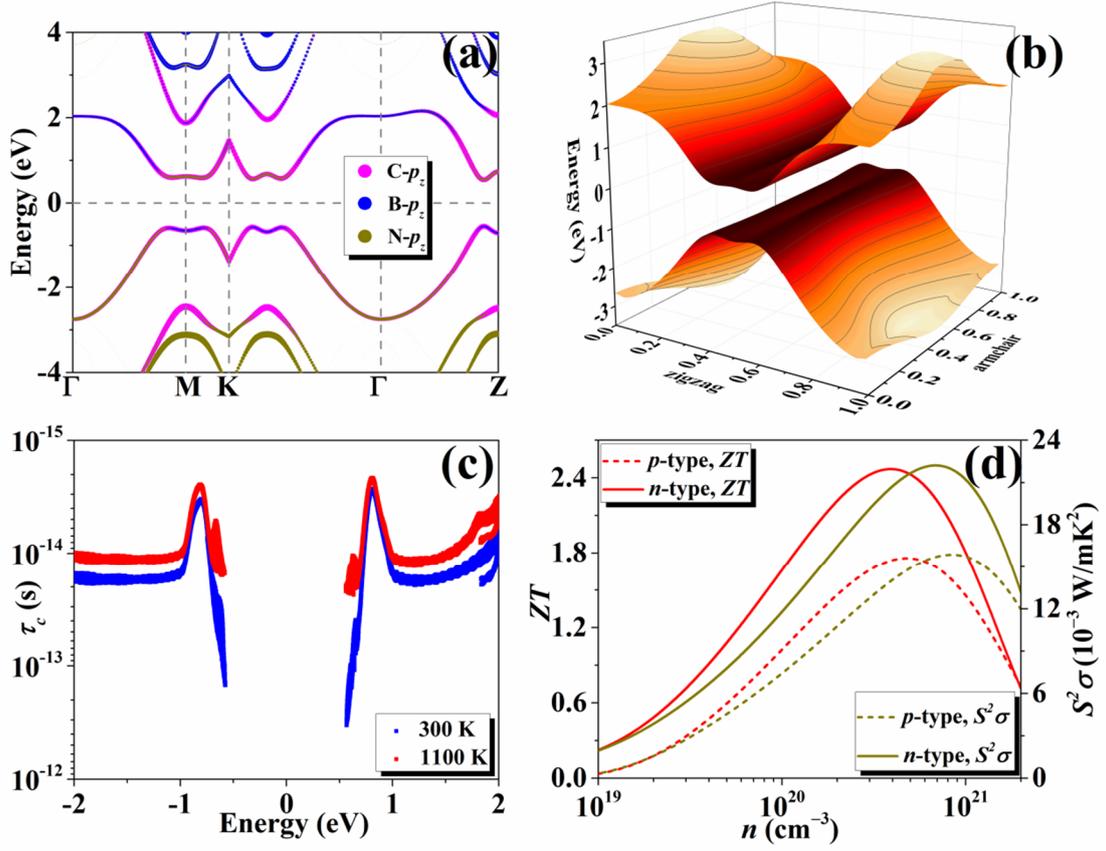

**Figure 2.** (a) Orbital-decomposed band structures of the graphene/h-BN superlattice monolayer. (b) The three-dimensional energy dispersion relations of the top valence band and bottom conduction band. (c) The energy-dependent carrier relaxation time of the superlattice at 300 and 1100 K. The Fermi level is at 0 eV. (d) The $ZT$ values and power factors of the superlattice, plotted as a function of carrier concentration along the zigzag direction at 1100 K.

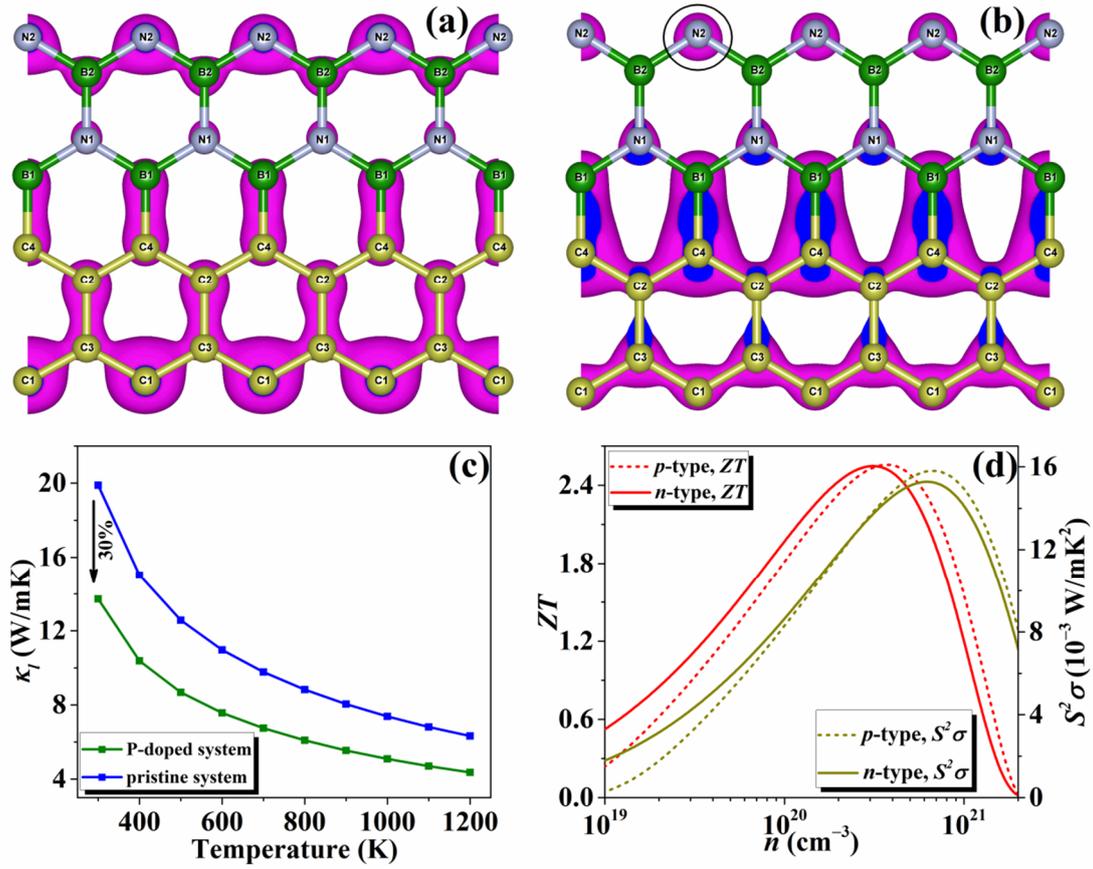

**Figure 3.** The band decomposed charge density of the graphene/h-BN superlattice monolayer around the (a) CBM and (b) VBM. (c) The lattice thermal conductivity of the supercell with nominal formula of $C_{24}B_{12}N_{11}P$, plotted along the zigzag direction as a function of temperature and compared with that of the pristine system. (d) The $ZT$ values and power factors of the P-substituted superlattice monolayer along the zigzag direction, plotted as a function of carrier concentration at 1100 K.

**References**

[1] L. D. Hicks and M. S. Dresselhaus, *Phys. Rev. B: Condens. Matter*, 1993, **47**, 12727–12731.

[2] L. D. Hicks and M. S. Dresselhaus, *Phys. Rev. B: Condens. Matter*, 1993, **47**, 16631–16634.

[3] M. S. Dresselhaus, G. Chen, M. Y. Tang, R. Yang, H. Lee, D. Wang, Z. Ren, J. P. Fleurial and P. Gogna, *Adv. Mater.*, 2007, **19**, 1043–1053.

[4] A. K. Geim and K. S. Novoselov, *Nat. Mater.*, 2007, **6**, 183–191.

[5] A. K. Geim, *Science*, 2009, **324**, 1530−1534.

[6] C. H. Liu, Y. C. Chang, T. B. Norris and Z. H. Zhong, *Nat. Nanotechnol.*, 2014, **9**, 273–278.

[7] A. A. Balandin, S. Ghosh, W. Bao, I. Calizo, D. Teweldebrhan, F. Miao and C, N, Lau, *Nano Lett.*, 2008, **8**, 902−907.

[8] A. H. Castro Neto, F. Guinea, N. M. R. Peres, K. S. Novoselov and A. K. Geim, *Rev. Mod. Phys.*, 2009, **81**, 109−162.

[9] P. J. Zomer, S. P. Dash, N Tombros and B. J. Van Wees, *Appl. Phys. Lett.*, 2011, **99**, 232104.

[10] C. Lee, X. Wei, J. W. Kysar and J. Hone, *Science*, 2008, **321**, 385–388.

[11] T. Gunst, T. Markussen, A.-P. Jauho and M. Brandbyge, *Phys. Rev. B*, 2011, **84**, 155449.

[12] Z.-X. Xie, L.-M. Tang, C.-N. Pan, K.-M. Li, K.-Q. Chen and W. Duan, *Appl. Phys. Lett.*, 2012, **100**, 073105.

[13] J. Y. Kim and J. C. Grossman, *Nano Lett.*, 2015, **15**, 2830–2835.

[14] R. Venkatasubramanian, E. Siivola, T. Colpitts and B. O'Quinn, *Nature*, 2001, **413**, 597−602.

[15] Z. liu, L. L. Ma, G. Shi, W. Zhou, Y. J. Gong, S. D. Lei, X. B. Yang, J. N. Zhang, J. J. Yu, K. P. Hackenberg, A. Babakhani, J.-C. Idrobo, R. Vajtai, J. Lou and P. M. Ajayan, *Nat. Nanotechnol.*, 2013, **8**, 119−124.

[16] J. Park, J. Lee, L. Liu, K. W. Clark, C. Durand, C. Park, B. G. Sumpter, A. P. Baddorf, A. Mohsin and M. Yoon, *Nat. Commun.*, 2014, **5**, 5403.

[17] R. Drost, A. Uppstu, F. Schulz, S. K. Hämäläinen, M. Ervasti, A. Harju and P. Liljeroth, *Nano Lett.*, 2014, **14**, 5128−5132.

[18] P. Karamanis, N. Otero and C. Pouchan, *J. Phys. Chem. C*, 2015, **119**,

11872−11885.

[19] Y. Hong, J. Zhang and X. C. Zeng, *Phys. Chem. Chem. Phys.*, 2016, **18**, 24164−24170.

[20] L. Zhu, R. Li and K. Yao, *Phys. Chem. Chem. Phys.*, 2017, **19**, 4085−4092.

[21] L. Ci, L. Song, C. Jin, D. Jariwala, D. Wu, Y. Li, A. Srivastava, Z. F. Wang, K. Storr, L. Balicas, F. Liu and P. M. Ajayan, *Nat. Mater.*, 2010, **9**, 430–435.

[22] P. Sutter, R. Cortes, J. Lahiri and E. Sutter, *Nano Lett.*, 2012, **12**, 4869–4874.

[23] V.-T. Tran, J. Saint-Martin and P. Dollfus, *Appl. Phys. Lett.*, 2014, **105**, 073114.

[24] Y. Ding, Y. Wang and J. Ni, *Appl. Phys. Lett.*, 2009, **95**, 123105.

[25] K. T. Lam, Y. Lu, Y. P. Feng and G. Liang, *Appl. Phys. Lett.*, 2011, **98**, 022101.

[26] J. Jung, Z. Qiao, Q. Niu and A. H. MacDonald, *Nano Lett.*, 2012, **12**, 2936–2940.

[27] Y. Yokomizo and J. Nakamura, *Appl. Phys. Lett.*, 2013, **103**, 113901.

[28] J.-W. Jiang, J.-S. Wang and B.-S. Wang, *Appl. Phys. Lett.*, 2011, **99**, 043109.

[29] H. Sevinçli, W. Li, N. Mingo, G. Guniberti and S. Roche, *Phys. Rev. B*, 2011, **84**, 205444.

[30] A. Kınacı, J. B. Haskins, C. Sevik, T. Çağın, *Phys. Rev. B*, 2012, **86**, 115410.

[31] T. Zhu and E. Ertekin, *Phys. Rev. B*, 2014, **90**, 195209.

[32] K. K. Yang, Y. P. Chen, R. D'Agosta, Y. Xie, J. X. Zhong and A. Rubio, *Phys. Rev. B*, 2012, **86**, 045425.

[33] S. I. Vishkayi, M. B. Tagani and H. R. Soleimani, *J. Phys. D: Appl. Phys.*, 2015, **48**, 235304.

[34] L. A. Algharagholy, Q. Algaliby, H. A. Marhoon, H. Sadeghi, H. M. Abduljalil and C. J. Lambert, *Nanotechnology*, 2015, **26**, 475401.

[35] V.-T. Tran, J. Saint-Martin and P. Dollfus, *Nanotechnology*, 2015, **26**, 495202.

[36] L. D. Zhao, G. Tan, S. Hao, J. He, Y. Pei, H. Chi, H. Wang, S. Gong, H. Xu and V. P. Dravid, *Science*, 2016, **351**, 141−144.

[37] W. Li, J. Carrete, N. A. Katcho and N. Mingo, *Comput. Phys. Commun.*, 2014, **185**, 1747 −1758.

[38] G. Kresse and J. Hafner, *Phys. Rev. B*, 1993, **47**, 558−561.

[39] G. Kresse and J. Hafner, *Phys. Rev. B*, 1994, **49**, 14251−14269.

[40] G. Kresse and J. Furthműller, *Comput. Mater. Sci.*, 1996, **6**, 15−50.

[41] A. Togo, F. Oba and I. Tanaka, *Phys. Rev. B*, 2008, **78**, 134106.

[42] P. Giannozzi, S. Baroni, N. Bonini, M. Calandra, R. Car, C. Cavazzoni, D. Ceresoli, *et al.*, *J. Phys.: Condens. Matter*, 2009, **21**, 395502.

[43] C. Hartwigsen, S. Goedecker and J. Hutter, *Phys. Rev. B*, 1998, **58**, 3641−3662.

[44] J. P. Perdew, K. Burke and M. Ernzerhof, *Phys. Rev. Lett.*, 1996, **77**, 3865−3868.

[45] J. Heyd, G. E. Scuseria and M. Ernzerhof, *J. Chem. Phys.*, 2006, **124**, 219906.

[46] K. Watanabe, T. Taniguchi, and H. Kanda, *Nat. Mater.*, 2004, **3**, 404−409.

[47] G. K. H. Madsen and D. J. Singh, *Comput. Phys. Commun.*, 2006, **175**, 67−71.

[48] J. Noffsinger, F. Giustino, B. D. Malone, C. H. Park, S. G. Louie and M. L. Cohen, *Comput. Phys. Commun.*, 2010, **181**, 2140−2148.

[49] F. Giustino, M. L. Cohen and S. G. Louie, *Phys. Rev. B*, 2007, **76**, 165108.

[50] S. Baroni, S. de Gironcoli, A. D. Corso and P. Giannozzi, *Rev. Mod. Phys.*, 2001, **73**, 515−562.

[51] Z. Liu, Y. J. Gong, W. Zhou, L. L. Ma, J. J. Yu, J. C. Idrobo, J. Jung, A. H. MacDonald, R. Vajtal, J. Lou and P. M. Ajayan, *Nat. Commun.*, 2013. **4**, 2541.

[52] G. Q. Ding, J. Carrete, W. Li, G. Y. Gao and K L Yao, *Appl. Phys. Lett.*, 2016, **108**, 233902.

[53] M. Hu, Y. H. Jing and X. L. Zhang, *Phys. Rev. B*, 2015, **91**, 155408.

[54] D. Wee, B. Kozinsky, N. Marzari and M. Fornari, *Phys. Rev. B*, 2010, **81**, 045204.

[55] L. Lindsay and D. A. Broido, *Phys. Rev. B*, 2011, **84**, 155421.

[56] D. J. Singh and I. I. Mazin, *Phys. Rev. B*, 1997, **56**, R1650−R1653.

[57] D. F. Zou, S. H. Xie, Y. Y. Liu, J. G. Lin and J. Y. Li, *J. Mater. Chem. A*, 2013, **1**, 8888−8896.

[58] Z. Z. Zhou, H. J. Liu, D. D. Fan, G. H. Cao and C. Y. Sheng, *ACS Appl. Mater. Interfaces*, 2018, **10**, 37031−37037.

[59] Y. Z. Pei, X. Y. Shi, A. LaLonde, H. Wang, L. D. Chen and J. Snyder, *Nature*, 2011, **473**, 66−69.

[60] J. Bardeen and W. Shockley, *Phys. Rev.*, 1950, **80**, 72−80.

[61] J. Park, Y. Xia and V. Ozoliņš, *Phys. Rev. Appl.*, 2019, **11**, 014058.

[62] D. Wu, L. D. Zhao, S. Q. Hao, Q. K. Jiang, F. S. Zheng, J. W. Doak, H. J. Wu, H. Chi, Y. Gelbstein, C. Uher, C. Wolverton, M. Kanatzidis and J. Q. He, *J. Am. Chem. Soc.*, 2014, **136**, 11412−11419.

[63] Y. F. Liu and P. F. P. Poudeu, *J. Mater. Chem. A*, 2015, **3**, 12507−12514.

[64] J. Q. Li, H. J. Wu, D. Wu, C. Y. Wang, Z. P. Zhang, Y. Li, F. S. Liu, W.-Q. Ao and J. Q. He, *Chem. Mater.*, 2016, **28**, 6367–6373.

[65] V. Martelli, J. L. Jiménez, M. Continentino, E. Baggio-Saitovitch and K. Behnia, *Phys. Rev. Lett.*, 2018, **120**, 125901.

[66] K. Lee, S. W. Kim, Y. Toda, S. Matsuishi and H. Hosono, *Nature*, 2013, **494**, 336–340.

[67] H. Q. Huang, K.-H. Jin, S. H. Zhang and F. Liu, *Nano Lett.*, 2018, **18**, 1972−1977.

[68] X. J. Ni, H. Q. Huang, K.-H. Jin, Z. F. Wang and F. Liu, arXiv:1904.03761.